\documentclass[runningheads]{llncs}
\usepackage{amsfonts}
\usepackage{graphicx}
\usepackage{amsfonts}
\usepackage{graphicx}
\usepackage{epsfig}
\usepackage{graphicx}
\usepackage{amsmath}
\usepackage{amssymb}
\usepackage{mathtools}
\usepackage{resizegather}
\usepackage{marvosym}
\usepackage{dsfont}
\usepackage{enumitem}
\usepackage{caption}
\usepackage{multirow}
\RequirePackage[utf8]{inputenc}
\usepackage{amsmath,bm}

\usepackage[colorlinks=true, allcolors=blue]{hyperref}

\usepackage[ruled, vlined]{algorithm2e}

\usepackage{lipsum}
\usepackage{array}
\usepackage{amsmath}

\usepackage{enumitem}
 
\usepackage{footnote}
\usepackage[hang, flushmargin]{footmisc}
\usepackage{footnotebackref}
\begin{document}
 
\title{Vicinal Feature Statistics Augmentation for Federated 3D Medical Volume Segmentation}
\titlerunning{Vicinal Data Augmented FL Segmentation}
% If the paper title is too long for the running head, you can set
% an abbreviated paper title here

\author{Yongsong Huang\inst{1,2} \and Wanqing Xie\inst{1,3} \and Mingzhen Li\inst{1,4} \and Mingmei Cheng\inst{1,2} \and Jinzhou Wu\inst{1}  \and Weixiao Wang\inst{1} \and Jane You\inst{5} \and Xiaofeng Liu\inst{1*}} 
 
\institute{Harvard Medical School, Harvard University, Boston, MA 02114, USA\and
Graduate School of
Engineering, Tohoku University, Sendai 980-8579, Japan\and Dept. of Intelligent Medical Engineering, School of Biomedical Engineering, Anhui Medical University, Hefei 230032, China\and Dept. of Mathematics, Washington University in St. Louis, MO 63130, USA\and
Dept. of Computing, The Hong Kong Polytechnic University, Hong Kong}

%$\dag$ contribute equally. 
 
\authorrunning{Y. Huang et al.}
\maketitle              % typeset the header of the contribution
 
\begin{abstract}

Federated learning (FL) enables multiple client medical institutes collaboratively train a deep learning (DL) model with privacy protection. However, the performance of FL can be constrained by the limited availability of labeled data in small institutes and the heterogeneous (i.e., non-i.i.d.) data distribution across institutes. Though data augmentation has been a proven technique to boost the generalization capabilities of conventional centralized DL as a "free lunch", its application in FL is largely underexplored. Notably, constrained by costly labeling, 3D medical segmentation generally relies on data augmentation. In this work, we aim to develop a vicinal feature-level data augmentation (VFDA) scheme to efficiently alleviate the local feature shift and facilitate collaborative training for privacy-aware FL segmentation. We take both the inner- and inter-institute divergence into consideration, without the need for cross-institute transfer of raw data or their mixup. Specifically, we exploit the batch-wise feature statistics (e.g., mean and standard deviation) in each institute to abstractly represent the discrepancy of data, and model each feature statistic probabilistically via a Gaussian prototype, with the mean corresponding to the original statistic and the variance quantifying the augmentation scope. From the vicinal risk minimization perspective, novel feature statistics can be drawn from the Gaussian distribution to fulfill augmentation. The variance is explicitly derived by the data bias in each individual institute and the underlying feature statistics characterized by all participating institutes. The added-on VFDA consistently yielded marked improvements over six advanced FL methods on both 3D brain tumor and cardiac segmentation.

%\keywords{Federated Learning, Data Augmentation, 3D Medical Segmentation, Feature Statistics, Vicinal Risk Minimization, MRI.} 

\end{abstract}

\section{Introduction}

Federated learning (FL)~\cite{rieke2020future} for medical image analysis, which follows a privacy-aware decentralized paradigm to learn a global model on several local medical institutes, has recently been the center of much attention. In practice, however, the participating institutes may have diverse data collection schemes, which can lead to inefficient global collaboration~\cite{rieke2020future,liu2022deep} in two ways. First, some of the small institutes may have limited training data to support effective local updating. This is especially evident in the medical segmentation task, which is often constrained by the availability of costly labeled training data~\cite{eaton2018improving}. Second, the data collected from different institutes with heterogeneous vendors, doses, and populations can result in biased gradient uploading to hinder the convergence of the federal model. Though in conventional centralized learning, the widely used data augmentation~\cite{shorten2019survey} can be a straightforward and unified solution to the challenges of limited sample and poor generalized data distribution~\cite{zhang2020generalizing}, it remains largely underexplored in FL segmentation.

It is of great importance to explore the proper global data augmentation scheme under the strict privacy restriction of FL, in which each client cannot access the data across institutes~\cite{rieke2020future}. Simply utilizing the data augmentation methods in centralized learning without injecting global information is sub-optimal, inheriting the institute's bias. Early attempts~\cite{li2019fedmd,lin2020ensemble,tuor2021overcoming} require accessing a global dataset to achieve balanced training, which is restricted in standard FL settings. Astraea~\cite{duan2020self} addresses this issue with several mediators between global servers and institutes. However, since each mediator group is a set of institutes, privacy among them is not protected. On the other hand, FedMix~\cite{yoon2021fedmix} and XORMix~\cite{shin2020xor} propose to transmit the averaged images across institutes, which still take risk of privacy breaches and are inherently weak at constructing semantic transform with image-level MixUp~\cite{liu2018data}. In addition, the above methods focus on classification and do not apply to segmentation--an important medical image analysis task with great demand for data augmentation~\cite{eaton2018improving}. 

%In addition, there are several efforts on improving the generalizability of the FL model from different perspectives by using dynamic regularizer~\cite{acar2021federated}, variance reduction~\cite{karimireddy2020scaffold}, local normalization~\cite{li2021fedbn}, and perturbed loss~\cite{qu2022generalized}. The federated data augmentation can be orthogonal to these methods and act as a general add-on solution to limited data scales and intrinsic distribution shifts. 
 
In this work, we propose to take both the inner- and inter-institute divergence into consideration without the need for any cross-institute transmission of raw data or their MixUp~\cite{yoon2021fedmix,shin2020xor}. To mitigate the aforementioned limitations, we resort to the feature-level vicinal risk minimization (VRM)~\cite{chapelle2000vicinal} to expand an example point in the feature space to a  probabilistically modeled flexible distribution, e.g., a conceptually simple Gaussian prototype, centered at the original point, with the variation reflecting the local and global data divergence. Novel feature statistics can be drawn from the Gaussian prototype distribution to fulfill augmentation. Moreover, inspired by the previous image style generation and domain adaptation works~\cite{chang2019domain,maria2017autodial,liu2022memory}, we propose to exploit the batch-wise feature statistics (e.g., mean and standard deviation) in each institute to abstractly represent the discrepancy of data, which are shared among institutes without any raw data transmission.

With our VFDA framework, the core issue is to properly associate the variance with the data bias in each individual institute and the underlying feature statistics characterized by all participating institutes. For effective augmentation, a proper variance is determined based on variances of feature statistics within each institute, regulated by the global variance of feature statistics characterized by all participating institutes. The augmentation in VFDA allows the local model to be trained over samples drawn from diverse feature distributions, alleviating local distribution shifts and facilitating institute-invariant representation learning, eventually leading to a better global model.

The main contributions of this work can be summarized as follows:

\noindent$\bullet$ To our knowledge, this is the first attempt to investigate an efficient data augmentation scheme in FL segmentation, which is especially pertinent for 3D medical data analysis.

\noindent$\bullet$ We propose a vicinal feature-level data augmentation (VFDA) scheme to model each feature statistic probabilistically via a conceptually simple Gaussian prototype and use the statistics drawn from the Gaussian distribution to implement augmentation. The transmission of raw images or their MixUp is not needed, meeting requirements for strict privacy protection. 

\noindent$\bullet$ Both local and global divergence are taken into account in the abstractly represented batch feature statistics to quantify the augmentation scope.  

\noindent$\bullet$ We evaluated our VFDA scheme on 3D brain tumor and cardiac anatomical segmentation tasks with 6 advanced FL methods to demonstrate its general efficacy and superiority.

\section{Methodology}

In FL segmentation, we are given $N$ local client institutes, in which the $n$-th institute has $M^n$ pairs of input volume $x_m^n\in\mathcal{X}$ and the corresponding segmentation label $y_m^n\in\mathcal{Y}$. We are expected to learn a global segmentation model ${f_g}({w_g}):\mathcal{X}\rightarrow\mathcal{Y}$ parameterized by ${w_g}$ across all institutes, generally enforced by the global empirical risk minimization (ERM) objectives:
\begin{align}
   \mathcal{L}_{ERM}^g({w_g}) = \frac{1}{N}\sum_{n=1}^N \mathbb{E}_{(x_m^n,y_m^n)\sim\mathcal{P}^n} [\mathcal{L}_{ERM}^n(x_m^N,y_m^N;{w_n})],\label{eq1}
\end{align} 
where $\mathcal{P}^n$ is the underlying distribution of institute $n$ (i.e., $\{x_m^n,y_m^n\}_{m=1}^{M^n}\sim\mathcal{P}^n$), and $\mathcal{L}_{ERM}^n$ is the institutes-wise empirical risk with local network parameter $w_n$. Specifically, in the segmentation task, $\mathcal{L}_{ERM}^n$ typically follows the empirical form of voxel-wise cross-entropy or Dice loss. With a privacy-aware decentralized setting, directly calculating $\mathcal{L}_{ERM}^g({w_g})$ with all data across institutes as conventional centralized training is infeasible. In contrast, FL~\cite{rieke2020future,mcmahan2017communication} relies on local training of models ${f_n}({w_n}), n\in\{1,\cdots,N\}$ in each institute in parallel with the local data only. In each round, the trained local models are aggregated to a global model ${f_g}({w_g})$, which is then further distributed to each institute for the next round of local training. Therefore, the local training objective in FL is equivalent to empirically approximating the local distribution $\mathcal{P}^n$ by a finite $M^n$ number of samples, i.e., $\mathcal{P}_{ERM}^n(x,y)=\frac{1}{M^n}\sum_{m=1}^{M^n}\delta(x=x_m^n,y=y_m^n)$, where $\delta(x=x_m^n,y=y_m^n)$ is a Dirac delta distribution with a point mass at $(x_m^n,y_m^n)$. 

Although the ERM objectives have been widely adopted in deep FL frameworks as the training scheme, the underlying assumption is that the empirical local distributions $\mathcal{P}_{ERM}^n$ are homologous to the underlying global distribution $\mathcal{P}^g$, which is unrealistic in actual clinical scenarios~\cite{qu2022generalized,liu2022memory}. In practice, there can be a significant performance drop since each $\mathcal{P}_{ERM}^n$ exhibits diverse data drifts with respect to $\mathcal{P}^n$ and $\mathcal{P}^g$, which lead to inconsistency of local and global empirical objectives and difficulties of generalization to testing distribution~\cite{acar2021federated,karimireddy2020scaffold}. In addition, each local model can be data starved, given only access to institute samples.  
 
\begin{figure}[t]
\begin{center} 
\includegraphics[width=1\linewidth]{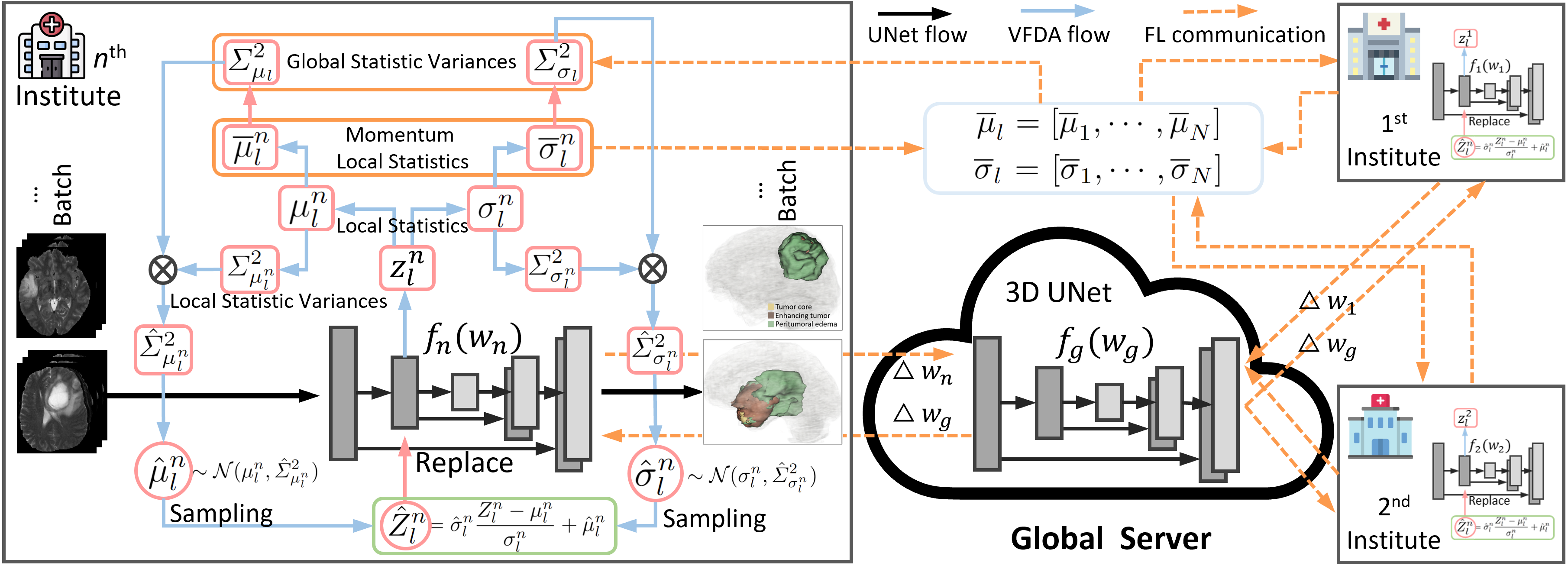}
\end{center} 
\caption{Illustration of our VFDA over an encoder layer in the classical FL 3D segmentation framework with heterogeneous clients.
}  
\label{fig1}\end{figure}

\subsection{Label Consistent Vicinal Feature Distribution Extrapolation}

Based on the above concerns, we propose to expand the Dirac delta distribution inherent in $\mathcal{P}_{ERM}^n(x,y)$ to a more expressive one to approximate the true distribution following the idea of VRM~\cite{chapelle2000vicinal,zhang2018mixup}. Therefore, we are able to examine the vicinal region of each sample $(x_m^n,y_m^n)$ for infinite data augmentation. Specifically, we can define a institute-wise vicinity distribution $\mathcal{V}^n(\hat{x}_m^n,\hat{y}_m^n|x_m^n,y_m^n)$ and apply it to  each sample $(x_m^n,y_m^n)$ to generate numerous pseudo samples $(\hat{x}_m^n,\hat{y}_m^n)$ and support the local institute training. More formally, we construct the vicinal local distribution as $\mathcal{P}_{VRM}^n=\frac{1}{M^n}\sum_{m=1}^{M^n}\mathcal{V}^n(\hat{x}_m^n,\hat{y}_m^n|x_m^n,y_m^n)$, and expect a better mimic of $\mathcal{P}^n$ with a proper $\mathcal{V}^n(\cdot)$.

There are numerous successful candidates of $\mathcal{V}^n(\cdot)$ in conventional 
centralized learning, e.g., MixUp~\cite{zhang2018mixup} and CutMix~\cite{yun2019cutmix}. Though simply utilizing MixUp~\cite{zhang2018mixup} or CutMix locally can potentially improve the performance (as shown in experiments), it can be sub-optimal as there are no global cues injected. In these cases, $\mathcal{P}_{VRM}^n$ gives a better mimic of the true local distribution $\mathcal{P}^n$ rather than the global distribution $\mathcal{P}^g$.  

Instead of the raw voxel level MixUp~\cite{zhang2018mixup,yun2019cutmix}, VFDA estimates a vicinity distribution $\mathcal{V}^n_l$ at each encoder layer $l$ to augment its batch-wise intermediate feature $Z^n_l$ in $n$-th institute for more flexible expansion~\cite{liu2018data}. Of note, $Z^n_l\in\mathbb{R}^{B\times C\times H\times W\times S}$ denotes the intermediate feature representation of $B$ mini-batch volumes, with height $H$, width $W$, slices $S$, and channels $C$. In addition, to achieve consistency training~\cite{xie2020unsupervised}, we define  $\mathcal{V}^n_l$ to be label consistent, i.e.,  $\mathcal{V}^n_l(\hat{Z}^n_l,\hat{Y}^n|Z^n_l,Y^n)=\mathcal{V}^n_l(\hat{Z}^n_l|Z^n_l)\delta(\hat{Y}^n={Y}^n)$, which only 
extrapolate the latent feature $Z^n_l$ while maintaining the consistency of the label ${Y}^n\in\mathbb{R}^{B\times H\times W\times S}$. Then, a key challenge is adaptively configuring $\mathcal{V}^n_l(\hat{Z}^n_l|Z^n_l)$ based on the local and global data divergence.

\subsection{Probabilistic Modeling of Feature Statistics}

As opposed to explicitly modeling $\mathcal{V}^n_l(\hat{Z}^n_l|Z^n_l)$, VFDA resorts to implicit feature augmentation by exploiting the batch-wise feature statistics in each institute to abstractly represent the discrepancy of input data, and model each feature statistic probabilistically via a Gaussian prototype. Specifically, we utilize the channel-wise feature statistics of mean $\mu^n_l$ and standard deviation $\sigma^n_l$. For $Z^n_l$, its channel-wise statistics of $\mu^n_l\in\mathbb{R}^{B\times C}$ and $\sigma^n_l\in\mathbb{R}^{B\times C}$ can be formulated as:
\begin{align}
   \mu^n_l&=\frac{1}{H\times W\times S}\sum_{h=1}^H\sum_{w=1}^W\sum_{s=1}^S Z^n_l;~~ \sigma^n_l=\sqrt{\frac{1}{H\times W\times S}\sum_{h=1}^H\sum_{w=1}^W\sum_{s=1}^S(Z^n_l-\mu^n_l)}.\label{eq2}
\end{align}
Recent works \cite{chang2019domain,maria2017autodial,wang2019transferable,mancini2018boosting} demonstrated that these low-order batch statistics are domain-specific, owing to the divergence of feature representations. As the abstraction of latent features, the feature statistics among local institutes will also exhibit inconsistency and follow shifts from the statistics of the true distribution. 

We propose to capture such shifts via probabilistic modeling. We hypothesize that each feature statistic follows a multi-variate Gaussian prototype, i.e., $\mu^n_l\sim\mathcal{N}(\mu^n_l,\hat{\Sigma}^2_{\mu^n_l})$ and $\sigma^n_l\sim\mathcal{N}(\sigma^n_l,\hat{\Sigma}^2_{\sigma^n_l})$, where each Gaussian prototype’s center corresponds to the original statistic, and the variance is an estimation of the potential feature statistic shift from the true distribution.

\subsection{Local and Global Statistic Variances Quantification}

Determining an appropriate variant range is challenging since each institute has only access to the data itself but has no sense of its statistical biases. Recent research~\cite{wang2019implicit,upchurch2017deep} has shown that deep feature space contains many semantic directions, and feature variances provide a reasonable measurement of potential meaningful semantic changes along the directions. This motivates us to estimate the variance of the Gaussian prototype from the variance of feature statistics.

\textbf{Local Statistic Variances}. In each institute, we compute local variances of feature statistics based on the information within each mini-batch:
\begin{align}
   {\Sigma}^2_{\mu^n_l}=\frac{1}{B}\sum_{b=1}^B(\mu^n_l-\mathbb{E}[\mu^n_l])^2\in\mathbb{R}^{C};~~ {\Sigma}^2_{\sigma^n_l}=\frac{1}{B}\sum_{b=1}^B(\sigma^n_l-\mathbb{E}[\sigma^n_l])^2\in\mathbb{R}^{C}, \label{eq3}
\end{align}
where ${\Sigma}^2_{\mu^n_l}$ and ${\Sigma}^2_{\sigma^n_l}$ denote the variance of feature mean $\mu^n_l$ and standard deviation $\sigma^n_l$ that are specific to each institute. Each value in ${\Sigma}^2_{\mu^n_l}$ and ${\Sigma}^2_{\sigma^n_l}$ is the variance of feature statistics in a particular channel, and its magnitude captures the potential change in that particular channel at that specific institute.

\textbf{Global Statistic Variances}. The institute-specific statistic variances are solely computed based on the data in each institute and are thus likely biased due to the bias in the local dataset. To resolve this, we further estimate the variances of the institute-sharing feature statistics, taking information from all institutes into account. Particularly, we propose a momentum version of feature statistics for each institute, which is updated online with an exponential momentum decay (EMD) strategy:
\begin{align}
   \overline{\mu}^n_l\leftarrow (1-\eta)\frac{1}{B}\sum_{b=1}^B {\mu}^n_l + \eta\overline{\mu}^n_l \in\mathbb{R}^{C};~~ \overline{\sigma}^n_l \leftarrow (1-\eta)\frac{1}{B}\sum_{b=1}^B {\sigma}^n_l + \eta\overline{\sigma}^n_l \in\mathbb{R}^{C},\label{eq4}
\end{align}
where the momentum factor $\eta=\eta^0\text{exp}(-r)$ follows an exponential decay over round $r$. Notably, $\overline{\mu}^n_l$ and $\overline{\sigma^n_l}$ are the momentum updated feature statistics of encoder layer $l$ in institute $n$. $\eta^0$ is empirically initialized to 10.

%Note that simply forcing the mean and variance in different domains to be the same can lead to a loss of expressiveness of the networks \cite{zhang2020generalizable}. 

In each communication round, these accumulated local feature statistics are sent to the server along with model parameters. Let $\overline{\mu}_l=[\overline{\mu}_1,\cdots,\overline{\mu}_N]\in\mathbb{R}^{N\times C}$ and $\overline{\sigma}_l=[\overline{\sigma}_1,\cdots,\overline{\sigma}_N]\in\mathbb{R}^{N\times C}$ denote the collections of accumulated feature statistics of all institutes, then the global, institute sharing statistic variances are calculated as:
\begin{align}
   {\Sigma}^2_{\mu_l}=\frac{1}{N}\sum_{n=1}^N(\overline{\mu}^n_l-\mathbb{E}[\overline{\mu}_l])^2\in\mathbb{R}^{C};~~ {\Sigma}^2_{\sigma_l}=\frac{1}{N}\sum_{n=1}^N(\overline{\sigma}^n_l-\mathbb{E}[\overline{\sigma}_l])^2\in\mathbb{R}^{C}. \label{eq5}
\end{align}
Along with the aggregated model parameters as in classical FL methods, e.g.,  FedAvg~\cite{mcmahan2017communication}, these variances are distributed back to each institute to inform a global estimation of feature statistic variances. Note that ${\Sigma}^2_{\mu_l}$ and ${\Sigma}^2_{\sigma_l}$ are shared by all participating institutes.

Institute sharing estimations ${\Sigma}^2_{\mu_l}$ and ${\Sigma}^2_{\sigma_l}$  provide a quantification of distribution divergence among institutes, and larger values imply potentials of more significant changes of the corresponding channels in the true feature statistic space. Therefore, for each institute, we weight the institute-specific statistic variances ${\Sigma}^2_{\mu^n_l}$, ${\Sigma}^2_{\sigma^n_l}$ with ${\Sigma}^2_{\mu_l}$, ${\Sigma}^2_{\sigma_l}$, so that each institute has a sense of such global divergence, i.e., $\hat{\Sigma}^2_{\mu^n_l}={\Sigma}^2_{\mu_l}{\Sigma}^2_{\mu^n_l}$ and $\hat{\Sigma}^2_{\sigma^n_l}={\Sigma}^2_{\sigma_l}{\Sigma}^2_{\sigma^n_l}$.

\subsection{Implementation of Vicinal Feature-level Data Augmentation}

After establishing the Gaussian prototype, we calculate novel feature $\hat{Z}^n_l$ in the vicinity of $Z^n_l$ as follows:
\begin{align}
   \hat{Z}^n_l=\hat{\sigma}^n_l\frac{{Z}^n_l-{\mu}^n_l}{{\sigma}^n_l}  +\hat{\mu}^n_l;~~ \hat{\mu}^n_l\sim\mathcal{N}(\mu^n_l,\hat{\Sigma}^2_{\mu^n_l}),~~ \hat{\sigma}^n_l\sim\mathcal{N}(\sigma^n_l,\hat{\Sigma}^2_{\sigma^n_l}),\label{eq6}
\end{align}
where ${Z}^n_l$ is first normalized with its original statistics by $\frac{{Z}^n_l-{\mu}^n_l}{{\sigma}^n_l}$, and further scaled with novel statistics $\hat{\mu}^n_l$ and $\hat{\sigma}^n_l$ that are randomly sampled from the corresponding Gaussian distribution~\cite{li2021feature}. To make the sampling differentiable, we apply the re-parameterization trick~\cite{kingma2013auto}:
\begin{align}
     \hat{\mu}^n_l=\mu^n_l+\epsilon_\mu\hat{\Sigma}^2_{\mu^n_l},~~ \hat{\sigma}^n_l=\sigma^n_l+\epsilon_\sigma\hat{\Sigma}^2_{\sigma^n_l}),\label{eq7}
\end{align}
where $\epsilon_\mu\sim\mathcal{N}(0,1)$ and $\epsilon_\sigma\sim\mathcal{N}(0,1)$ follow normal distribution. 

The proposed VFDA in Eq.(\ref{eq6}) works in a plug-and-play fashion, which can be inserted at arbitrary positions of the model to facilitate latent semantic augmentation. In our implementation, we add a VFDA after each encoder layer of UNet. Of note, we explore the batch-wise feature statistics for vicinal expansion while not relying on the network with batch normalization layer~\cite{liu2022memory,li2021fedbn}. Our FDA can be widely generalized to modern deep learning models with mini-batch training. During testing, no augmentation is performed.

\section{Experiments and Results}

To show the effectiveness of our framework, we experimented on both 3D federated brain tumor and cardiac anatomical segmentation tasks and added VFDA to numerous advanced FL segmentation models, e.g., FedAvg~\cite{mcmahan2017communication}, FedProx~\cite{gudur2021resource}, FedBN~\cite{li2021fedbn}, FedNorm~\cite{yin2022efficient}, PRRF~\cite{chen2021personalized}, and FedCRLD~\cite{qi2022contrastive}. Of note, our VFDA can be directly applied to 2D segmentation by setting $S=0$. 

We implemented all modules on a server with an NVIDIA A100 GPU and used the PyTorch toolbox. For the evaluation metrics, we employed the widely accepted Dice similarity coefficient (DSC), which measures the overlap between the predicted segmentation mask and the label.

\begin{table*}[t]
\centering
\caption{Comparisons of different FL methods with or without our VFDA framework in FeTS2021, and the ablation study of VFDA modules.}\vspace{+5pt}
\resizebox{1\linewidth}{!}{
\begin{tabular}{c|c|cccc}
\hline

&Additional& & Dice & Score & [$\%$] $\uparrow$  \\ \cline{3-6}

{Method}& augmentation& ~~ET~~ & ~~TC~~ & ~~WT~~ & ~~Mean~~ \\ \hline\hline

FedAvg~\cite{mcmahan2017communication}  & - & 75.63 &62.01 &76.55& 71.14\\    
FedAvg~\cite{mcmahan2017communication}  & MixUp & 75.80 &62.74 &76.95& 71.83\\      \hline  
FedAvg~\cite{mcmahan2017communication}  & VFDA (Ours) & \textbf{76.59} &\textbf{64.10} &\textbf{77.86}& \textbf{72.85}\\ 
FedAvg~\cite{mcmahan2017communication}  &  VFDA w/o EMD &76.08 &63.47 &77.29& 72.28\\ 

FedAvg~\cite{mcmahan2017communication}  &  VFDA w/o Global Statistic Variances& 75.92&63.15 &77.03& 72.03\\ \hline  \hline

FedNorm~\cite{yin2022efficient} & - & 75.60 &63.76& 76.98& 72.11 \\  
FedNorm~\cite{yin2022efficient} & MixUp & 75.82 &63.96& 77.05& 72.28  \\  \hline
FedNorm~\cite{yin2022efficient} & VFDA (Ours)  & \textbf{76.49} &\textbf{64.82} &\textbf{78.17}& \textbf{73.16}\\
FedNorm~\cite{yin2022efficient} & VFDA w/o EMD & 76.14&64.55& 77.83& 72.84  \\  
FedNorm~\cite{yin2022efficient} & VFDA w/o Global Statistic Variances& 76.01 &64.34& 77.50& 72.62  \\ \hline
\end{tabular}}\label{tab1}
\end{table*}

\subsection{Federated Brain Tumor Segmentation}
 
The Federated Tumor Segmentation (FeTS) 2021 Challenge Task-1~\cite{pati2021federated} incorporates the magnetic resonance imaging (MRI) volumes and segmentation label of three brain tumor structures, i.e., tumor core (TC), enhanced tumor (ET) and whole tumor (WT), from 341 subjects. For FL, we followed the standard dataset partition-2 to separate 22 local clients based on their original institutions and further split the large institutes into subsets according to the tumor sizes, which are subject-independent. MRI scans can be highly heterogeneous among participating clients in the FeTS challenge as various scanners and image protocols were employed. The size-based subset also involves the divergence of different tumor grades. With the fine-grained partition in FeTS partition-2, the subjects in each local client are limited. 
 
\begin{figure}[t]
\begin{center}
\includegraphics[width=1\linewidth]{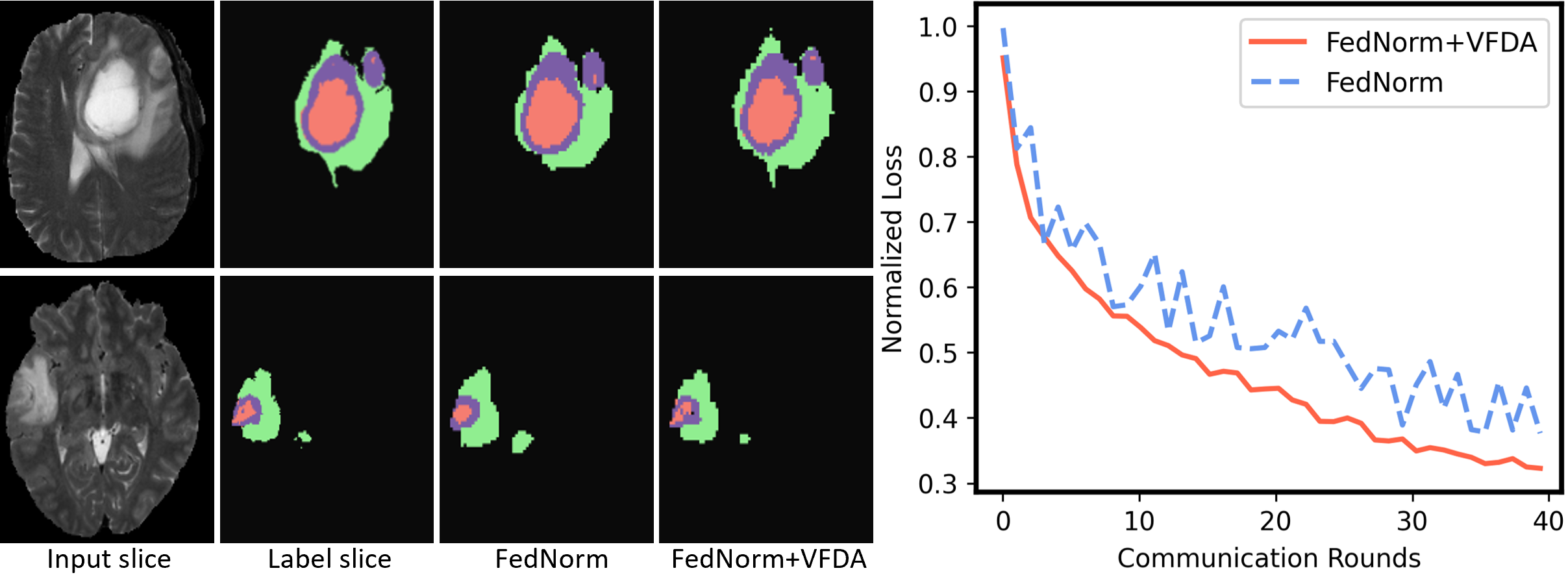}
\end{center} \vspace{-5pt}
\caption{Left: The qualitative comparisons of 2 slices from two subjects with FedNorm with or without our VFDA. CT: red; ET: purple; WT: red+purple+green; background: black (best viewed in color). Right: the normalized training loss of FedNorm with or without our VFDA.}  
\label{exp1} 
\end{figure}

According to the standard evaluation protocol, the segmentation model of ResNet-based 3D UNet is fixed for all participants~\cite{pati2021federated}. We adopted the successful solutions in FeTS, i.e., FedNorm~\cite{yin2022efficient} and FedAvg~\cite{mcmahan2017communication} with different model aggregation schemes as our baselines. For fair comparisons, we followed the detailed setting of federated aggregation with tensor normalization (FedNorm)~\cite{yin2022efficient} based on the open-fl framework. All of the experiments were conducted under a fixed train validation split and random seed to make our results convincing and deterministic. Specifically, in each communication round, we performed one epoch for local client training and initialized the learning rate to 5e-4 with a polynomial decaying factor of 0.9 consistently. For all methods, we added the vanilla data augmentation of rotation, scaling, elastic deformation, brightness, and aggressiveness adjustment as in~\cite{yin2022efficient}. 

The quantitative evaluation results with respect to Dice score are presented in Table~\ref{tab1}. Simply adding the MixUp augmentation improved the performance of both FedAvg and FedNorm significantly. By taking a more flexible feature-level expansion based on the local and global divergence described by the abstract feature statistic variances, our VFDA can achieve remarkable results. For the ablation study, we removed the global statistic variances or EMD module and denoted them as VFDA w/o EMD or  VFDA w/o Global Statistic Variances, respectively. Their inferior performance compared to VFDA demonstrates the contribution of global statistic variances or EMD training.

In Fig.\ref{exp1}, some example slices are shown, in which the VFDA achieves more accurate delineations compared to FedNorm. In addition, adding VFDA contributes to a stabler continuous optimization process than FedNorm.

%We also analyzed the additional communication cost of VFDA over the vanilla FedAvg~\cite{mcmahan2017communication}. In each round, we transmited the averaged feature statistics $\overline{\mu}^n_l\in\mathbb{R}^C$ and $\overline{\sigma}^n_l\in\mathbb{R}^C, \forall n, l$ from local institutes to global server, and ${\Sigma}^2_{\mu_l}\in\mathbb{R}^C$ and ${\Sigma}^2_{\sigma_l}\in\mathbb{R}^C, \forall l$ from global server to local institutes. Thus, for each encoder layer, VFDA required transmitting 4$C$×32 bit (i.e., 16$C$ Byte) additional data in each round. For our UNet model, the total additional communication cost was approximately 31 KB\footnote{[16 × (64 + 128 + 256 + 512 + 1024)]/1024 = 31 KB}, which was negligible compared with the dual-direction transmission of typical UNet models with at least a few MB of parameters. 

\begin{table*}[t]
\centering
\caption{Comparisons of different FL methods with or without our VFDA framework in cardiac segmentation task, and the ablation study of VFDA modules.}\vspace{+5pt}
\resizebox{1\linewidth}{!}{
\begin{tabular}{c|c|ccccccc}
\hline

&Additional& & && Dice & Score & [$\%$] $\uparrow$  \\ \cline{3-9}

{Method}& augmentation& ~~~A~~~ & ~~~B~~~ & ~~~C~~~ & ~~~D~~~ & ~~~E~~~ & ~~~F~~~  & ~~~Mean~~~ \\ \hline\hline

FedAvg~\cite{mcmahan2017communication}  & - & 85.84 &85.39 &89.08& 79.77 &84.42& 18.36&73.81\\
FedAvg~\cite{mcmahan2017communication}  &  VFDA w/o Global  & 87.42 &86.25&88.30&82.59 &85.45& 45.08&79.18\\  
FedAvg~\cite{mcmahan2017communication}  &  VFDA (Ours)  & \textbf{88.72} &\textbf{87.53} &\textbf{89.91}& \textbf{84.15} &\textbf{86.05}& \textbf{52.47}& \textbf{81.4} \\ \hline  \hline

FedProx~\cite{gudur2021resource}  & - &86.70 &84.41&88.81& 82.85 &83.66& 27.26&75.62\\
FedProx~\cite{gudur2021resource}  &  VFDA w/o Global   & 87.65 &86.40 &89.28& 84.91 &85.13& 49.72&80.51\\
FedProx~\cite{gudur2021resource}  &  VFDA (Ours)   & \textbf{88.23} &\textbf{87.13} &\textbf{89.95}& \textbf{84.86} &\textbf{85.60}& \textbf{52.81}&\textbf{81.43}\\ \hline  \hline

FedBN~\cite{li2021fedbn}  & - & 86.98 &85.87 &89.58& 81.91 &84.73& 27.49&76.09\\    
FedBN~\cite{li2021fedbn}  &  VFDA w/o Global   & 88.05 &86.85 &89.74& 83.62 &85.82& 45.65&79.96 \\
FedBN~\cite{li2021fedbn}  &   VFDA (Ours)   & \textbf{88.62} &\textbf{87.28} &\textbf{90.20}&\textbf{85.07} &\textbf{86.15}& \textbf{48.30}&\textbf{80.94} \\ \hline  \hline

PRRF~\cite{chen2021personalized}  & - & 87.04 &86.11 &88.05& 84.65 &83.85& 53.09&80.47\\    
PRRF~\cite{chen2021personalized}  &  VFDA w/o Global   & 87.86 &87.03 &89.65& 85.12 &85.44& 61.05&82.69 \\
PRRF~\cite{chen2021personalized}  &  VFDA (Ours)   & \textbf{88.92} &\textbf{88.75} &\textbf{90.26}& \textbf{86.07}&\textbf{86.29}& \textbf{62.38}&\textbf{83.79} \\ \hline  \hline

FedCRLD~\cite{qi2022contrastive}~  & - & 88.06 &87.28 &90.88& 86.96 &86.40& 76.15&85.96\\ 
FedCRLD~\cite{qi2022contrastive}~  &  ~VFDA w/o Global~   & 89.02&88.34 &91.24& 87.35 &87.23& 77.80&86.83 \\
FedCRLD~\cite{qi2022contrastive}~  &  VFDA (Ours)   & \textbf{89.74} &\textbf{89.05} &\textbf{92.11}& \textbf{87.83}&\textbf{87.62}& \textbf{78.53}& \textbf{87.48}\\ \hline  
\end{tabular}}\label{tab2}
\end{table*}

\subsection{Federated Cardiac Anatomical Segmentation}

To further demonstrate the generalizability of VFDA, we also evaluated it on multi-center multi-sequence cardiac MRI segmentation as in~\cite{qi2022contrastive}. Specifically, a real-world FL task is constructed using the publicly available M\&M \cite{campello2021multi} and Emidec \cite{lalande2020emidec} datasets. Of note, M\&M dataset incorporates the cine-MRI of the subjects from five centers in three countries and is scanned with four different scanner vendors, and the Emidec is a 
delayed enhancement (DE) MRI dataset. As in~\cite{qi2022contrastive}, the M\&M dataset is split into five institutes/centers, i.e., client A-E, while Emidec is configured as the sixth institute, i.e., client F. There are notable appearance shifts across institute due to the diverge centers, devices, and contrast agents. We chose the subject-independent 7/1/2 split for each client institute and adopted the 3D UNet as the segmentation model backbone.

\begin{figure}[t!]
\begin{center}
\includegraphics[width=1\linewidth]{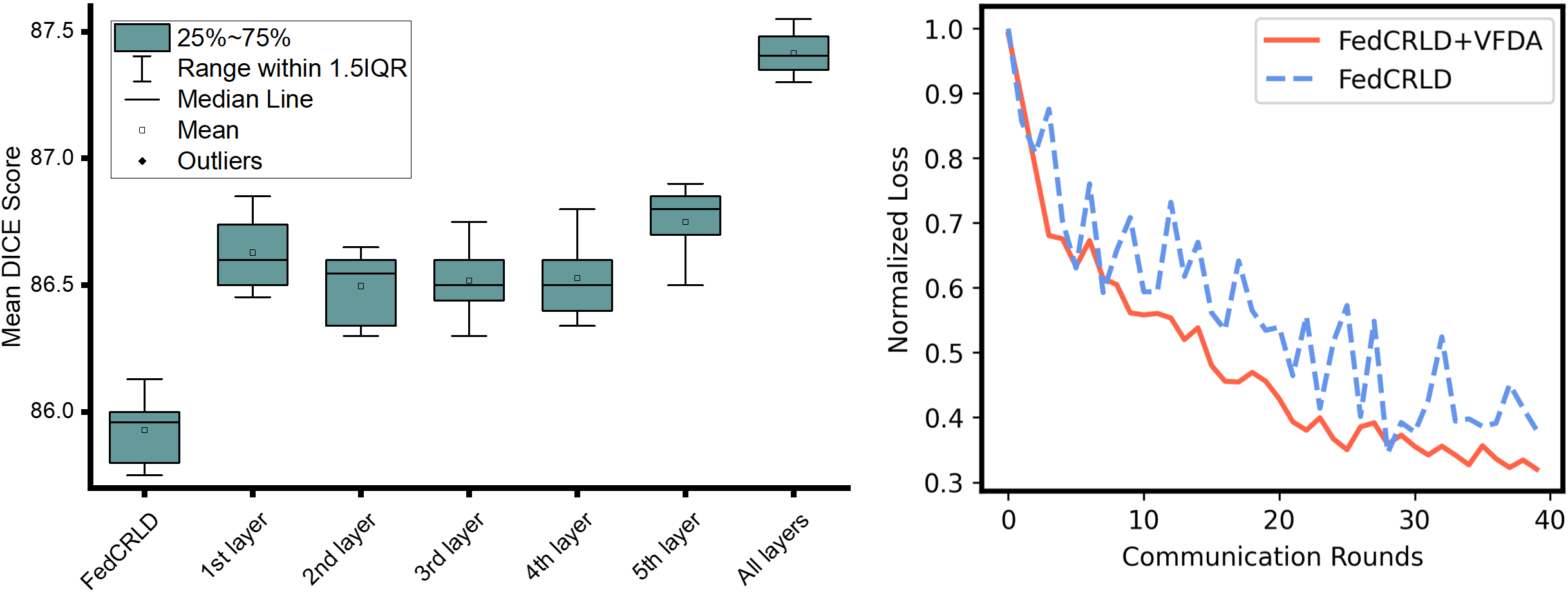}
\end{center} \vspace{-5pt}
\caption{Left: Sensitivity analysis of adding VFDA to each layer of the 3DUNet encoder in FedCRLD with five times random trails. Right: the normalized training loss of
FedCRLD with or without our VFDA.}  
\label{exp2}
\end{figure}

For fair comparisons, we adopted the vanilla augmentation of rotation, translation,
scale, and mirror as~\cite{qi2022contrastive}. Notably, the compared baselines of FedProx~\cite{gudur2021resource}, FedBN~\cite{li2021fedbn},  PRRF~\cite{chen2021personalized}, and FedCRLD~\cite{qi2022contrastive} are designed for non-IID cases, which explicitly target the data bias among local institutes using different techniques. However, how to efficiently induce the global data divergence for FL debias is a long-lasting challenge. Our VFDA can be a unified and general solution to simply add on these methods to efficiently improve the segmentation performance, as shown in Table~\ref{tab2}. In the ablation study, we also demonstrated that taking the global statistic variances into consideration is important for VFDA for all these baselines.

In Fig.\ref{exp2} left, we investigated the effect of VFDA in different encoder layers. Of note, adding VFDA to the decoder layer does not improve the performance. In Fig.\ref{exp2} right, the FedCRLD training in client F (i.e., Emidec with different DE-MRI) is relatively unstable, while adding VFDA smooths the optimization and leads to a lower loss.

\section{Conclusion}
 
In this work, we proposed a novel and efficient data augmentation methodology for federated learning in 3D medical volume segmentation, which suffers from the imbalance of clients in small institutes, and the inner- and inter-institute heterogeneous data shift. We resort to batch-wise feature statistics as an abstract quantification of the local and global statistic variances and utilize them probabilistically via a Gaussian prototype. We utilize the mean corresponding to the original statistic and the variance to define the proper augmentation scope in a label-preserving vicinal risk minimization framework on the feature space to expand the feature by simply sampling the Gaussian distribution with the re-parameterization trick. The experiments in both 3D brain tumor and cardiac anatomical structure segmentation FL tasks with six advanced FL methods consistently demonstrate its efficiency and generality. It has the potential to be widely adapted to different FL scenarios with low additional community costs.

\bibliographystyle{splncs04}
\bibliography{egbib}

\end{document}